\documentclass[%
reprint,
superscriptaddress,
amsmath,
amssymb,
aps,
pra,
]{revtex4-1}

\usepackage{graphicx}% Include figure files
\usepackage{dcolumn}% Align table columns on decimal point
\usepackage{bm}% bold math
\usepackage{color}
\usepackage[T1]{fontenc}
\usepackage{mathptmx}
\usepackage{etoolbox}
\usepackage{siunitx}
\usepackage{longtable}
\usepackage{wrapfig}
\usepackage{rotating}
\usepackage[normalem]{ulem}
\usepackage{amsmath}
\usepackage{amssymb}
\usepackage{capt-of}
\usepackage[english]{babel}
\usepackage{amsfonts}
\usepackage{braket}
\usepackage{tensor}
\usepackage{standalone}
\usepackage{siunitx}
\usepackage{tikz}
\usepackage{standalone}
\usepackage{enumitem}

\setlist[itemize]{leftmargin=*}
\DeclareMathOperator{\1}{\mathbb{1}}

\begin{document}
\title{Strain-enabled control of the vanadium qudit in silicon carbide}

\author{Philipp Koller}
\affiliation{Institute for Quantum Optics and Quantum Information (IQOQI), Austrian Academy of Sciences, Boltzmanngasse 3, 1090 Vienna, Austria}
\affiliation{University of Vienna, Faculty of Physics \& Vienna Doctoral School in Physics,  Boltzmanngasse 5, A-1090 Vienna, Austria}
\author{Thomas Astner}
\affiliation{Institute for Quantum Optics and Quantum Information (IQOQI), Austrian Academy of Sciences, Boltzmanngasse 3, 1090 Vienna, Austria}
\author{Benedikt Tissot}
\affiliation{Department of Physics, University of Konstanz, D-78457 Konstanz, Germany}
\affiliation{Center for Hybrid Quantum Networks, Niels Bohr Institute, University of Copenhagen, Blegdamsvej 17, 2100 Copenhagen Ø, Denmark}
\author{Guido Burkard}
\affiliation{Department of Physics, University of Konstanz, D-78457 Konstanz, Germany}
\author{Michael Trupke}
\email{michael.trupke@oeaw.ac.at}
\affiliation{Institute for Quantum Optics and Quantum Information (IQOQI), Austrian Academy of Sciences, Boltzmanngasse 3, 1090 Vienna, Austria}

\date{\today}% It is always \today, today,
             %  but any date may be explicitly specified

\begin{abstract}
Vanadium in silicon carbide is a promising spin photon interface candidate with optical transitions in the telecom range and a long lived electron spin, hosted in an advanced semiconductor platform.
In this detailed investigation of the defect's 16-dimensional ground state spin manifold at millikelvin temperatures, a wide range of previously unreported transitions are observed which are accurately described using a theoretical model that includes strain.
Using a superconducting microcoil we achieve Rabi frequencies exceeding \SI{20}{\mega\hertz} and perform the first coherent manipulation of a direct hyperfine transition.
These insights further underscore the defect's potential for strain engineering and sensing, as well as for fault-tolerant qudit encoding.
\end{abstract}

\maketitle

\section{Introduction}
\label{sec:intro}

Semiconductor materials host a wide variety of optically active spin centres with numerous applications in classical and quantum technologies including sensing, photonics, and fundamental physics \cite{Awschalom2018,Doherty2013,Castelletto2022,Wolfowicz2021, Fernandez23, Yu2024, Vaartjes2024}. 
Vanadium in silicon carbide has recently been found to possess several attractive features for quantum photonics and quantum communication.
Its neutral charge state (V$^{4+}$) has relatively fast, spin-dependent optical transitions in the telecom range, with lifetimes between 10 ns and 170 ns, and extremely small inhomogeneous broadening in isotopically purified material \cite{Spindlberger2019,Wolfowicz2020,Cilibrizzi2023}.
In 4H SiC, vanadium has been detected in two different sites, termed $\alpha$ and $\beta$.
The $\alpha$ defect studied in this work presents strong luminescence at $\SI{1.28}{\micro\metre}$ and has a long  $S=1/2$ spin relaxation lifetime in excess of \SI{20}{\second} at \SI{100}{\milli\kelvin} \cite{Astner2024, Ahn2024}.
Magnetic resonance measurements have revealed long spin dephasing times $T^*_2$ on the order of microseconds and a spin coherence lifetime $\gg$\SI{20}{\micro\second} at a temperature of \SI{2}{\kelvin} \cite{Hendriks2022}. The $I=7/2$ nuclear spin may allow error protection via qudit encoding \cite{Lim2023}. However, many of these features, including the spin properties at ultralow temperatures and the effects of strain on the optical and spin properties of the system, require detailed scrutiny in order to ascertain its suitability for quantum computing and communication \cite{Tissot2024, Nemoto2014, Ecker2024}.

\begin{figure*}
    %\centering
    \includegraphics[scale=1, page=1]{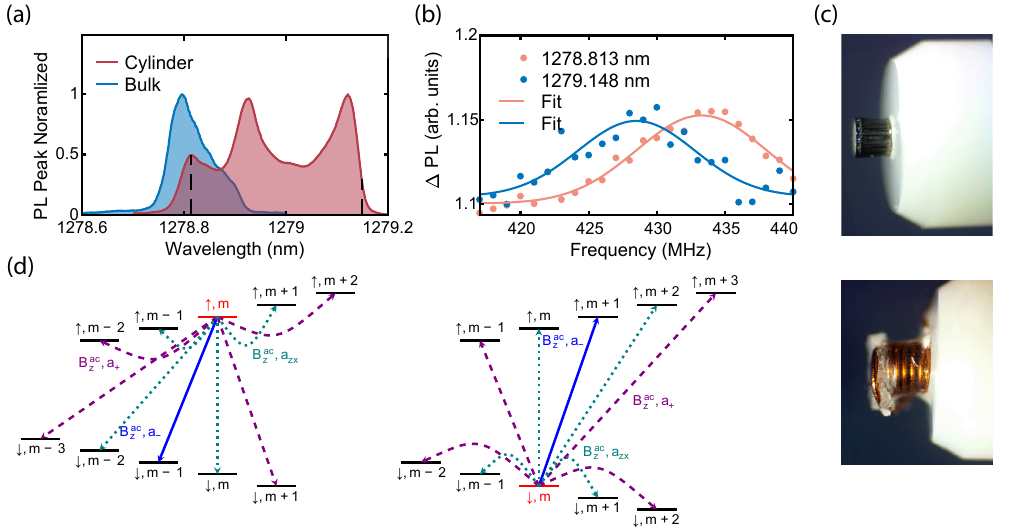}
    \caption{\label{fig:PLEODMRHAMPIC}
    (a) PLE spectra of V$^{4+}$ $\alpha$ ensembles. Blue: ensemble in a bulk, \SI{5}{\milli\meter}$\times$\SI{5}{\milli\meter}$\times$\SI{492}{\micro\meter} 4H SiC chip. Red: micromachined SiC cylinder.
    (b) ODMR measurements in the SiC cylinder of the lowest clock transition in the ground-state manifold at \SI{29}{\milli\tesla} at two different wavelengths, also indicated by dashed lines in panel (a). 
    The solid lines are Gaussian fits to the two measurements, yielding central frequencies of \SI{433.3 \pm 0.5}{\mega\hertz} and \SI{428.5 \pm 0.8}{\mega\hertz} at excitation wavelengths of \SI{1278.813}{\nano\meter} and \SI{1279.148}{\nano\meter}, respectively.
    (c) Images of the sample. A cylindrical SiC crystal with a diameter of \SI{290}{\micro\meter} is embedded in a SC microcoil at the tip of an optical single-mode fiber. A macroscopic SC coil is used to apply on-axis magnetic bias fields.
    (d) Modification of selection rules for transitions from an initial state, marked in red, in the presence of small transverse strain. Blue: Allowed transitions in the strain-free system. Green, purple: strain-induced transitions with linear strain dependence of the transition matrix element.
    }
\end{figure*}

In this work, we investigate the 16-dimensional spin manifold on a V$^{4+}$ $\alpha$ ensemble in 4H SiC extensively using strong radio-frequency driving fields, while maintaining ultralow temperatures on the order of \SI{100}{\milli\kelvin}. 
Our sample, a micromachined SiC crystal, presents a significantly broadened PLE trace compared to bulk material (see Fig.~\ref{fig:PLEODMRHAMPIC}).
In addition to the expected ground state spin resonances, we observe features which correspond to direct hyperfine transitions within one spin manifold.
A likely explanation of the observed differences to previous reports is given by the presence of strain in the sample, which leads to a modification of the spin basis \cite{Tissot2024,Guo2023}. 
In order to understand these effects in detail, we build upon very recent theoretical work \cite{Tissot2024} and perform targeted measurements on the spin manifold..
We furthermore measure Rabi and Ramsey spectra of a clock transition in this manifold, providing further insights into the spin properties of the system.

We now adapt the formalism developed in \cite{Tissot2024} to investigate how small strain affects the hyperfine interaction.
The allowed transitions for parallel driving are determined by the selection rules and can be understood in terms of conservation of angular momentum.
Since the V$^{4+}~\alpha$ defect ground state is a Kramers doublet (KD), the total angular momentum of the pseudospin $|\uparrow\rangle,\, |\downarrow\rangle$ states is the sum of their electron spin and angular momentum components. 
To leading order, these are given by $m_s=\mp 1/2$ and an orbital contribution which transforms like $m_l=\pm1$, respectively.
In the strain-free system $|\downarrow,m\rangle \leftrightarrow |\uparrow,m+1\rangle$ transitions are therefore allowed \cite{Tissot2021spin,Tissot2021hyper,Tissot2022}. 
The application of strain causes mixing of the states, leading to further allowed transitions.
The Hamiltonian describing the KD ground state under small strain and for a magnetic field \(B_{\mathrm{tot}}(t) = B + B_{\mathrm{ac}} f(t)\) along the crystal axis (\(\vec{e}_z\)) is
  \begin{equation}
  \label{eq:Hg}
  \begin{split}
  H_g =\, &(\mu_B g_z S_z + \mu_N g_N I_z) B_{\mathrm{tot}}(t) + a_{zz} S_z I_z\\
  &+a_{-} (S_x I_x - S_y I_y) + a_{zx} S_z I_x \\
  &+ a_{+} (S_x I_x + S_y I_y) + a_{xz} S_x I_z ,
  \end{split}
  \end{equation}
with the pseudospin-1/2 operator for the KD \(\vec{S}\) and the nuclear spin \(\vec{I}\) in units of \(\hbar\), the Bohr (nuclear) magneton \(\mu_B\) ($\mu_N$), the parallel (nuclear) g-factor of the KD \(g_z\) ($g_N$), as well as the hyperfine tensor constants \(a_{ij}\) (\(i,j=x,y,z\))  \cite{Tissot2024}.
Here, $2a_{xx}=a_{+}+a_{-}$ and $2a_{yy}=a_{+}-a_{-}$.
The components \(a_{+}\) and \(a_{zx}\) are linear in strain to leading order, while the leading contribution for \(a_{xz}\) is quadratic in strain.
The remaining parameters \(g_z,\,\, a_{zz},\,\, a_{-}\) have a strain-independent contribution and a strain-dependent correction with a quadratic leading order.

To interpret the allowed transitions we treat the additional hyperfine tensor elements caused by strain as a small perturbation, which is a good approximation for \(\mu_B g_z B \gg a_{ij}\) with \(i,j = x,y,z\) (additional information can be found in the Supplementary Material section III at \cite{SupplMat2024}).
Using perturbation theory we can assign the transitions allowed by an alternating magnetic field along the crystal axis driving the defect.
We show the resulting modification of the selection rules in Fig.~\ref{fig:PLEODMRHAMPIC}~(d).
The transitions allowed in perfectly strain-free samples (solid arrows) and those which have a linear dependence of their transition matrix elements on the strain (dotted and dashed arrows) are shown.

\begin{figure*}[t]
    \includegraphics[scale=1, page=3]{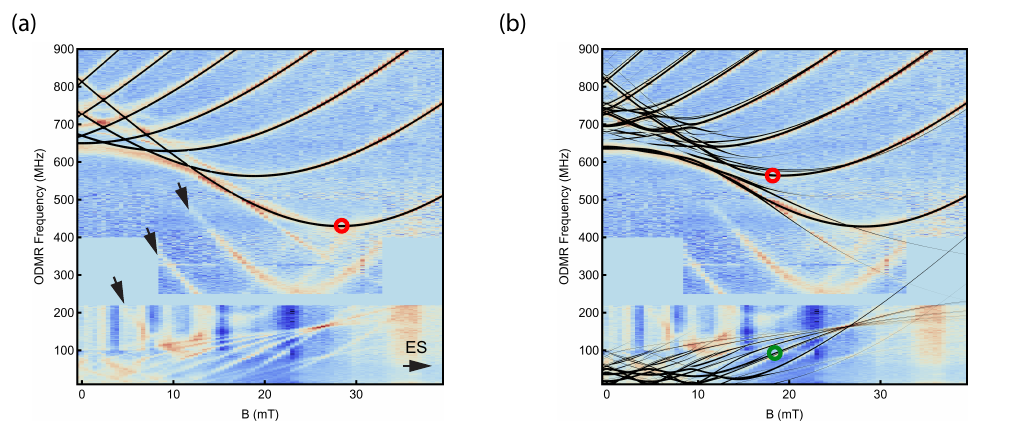}
    \caption{\label{fig:map} ODMR spectrum of the defect ensemble, recorded with an excitation wavelength of \SI{1279.12}{\nano\meter}. 
    The spectrum was recorded in different sets of measurements with different power settings, then normalized and stitched.
    (a) The dataset is shown with the calculated transitions of a strain-free V$^{4+} \alpha$ defect (black lines). 
    Black arrows indicate signals from spin transitions in the optically excited state.
    The red circle indicates the transition measured in Fig.~\ref{fig:PLEODMRHAMPIC}~(b).
    (b) The same measurement, but with a calculated transition spectrum using the strain-modified Hamiltonian (black lines).
    The red circle indicates the clock transition at which the measurements in Fig.~\ref{fig:rabi} (a)-(e) were taken.
    The green circle marks the hyperfine transition measured in Fig.~\ref{fig:rabi} (f). 
    The linewidth used to display the calculated spectra is proportional to the calculated signal strength in both figures.
    }
\end{figure*}
To better understand the V$^{4+}~\alpha$ defect ground state properties, we investigate an ensemble hosted in a 4H SiC microcrystal at millikelvin temperatures.
Using laser machining, a bulk crystal is cut to a cylindrical shape with diameter of \SI{290}{\micro\meter} and a height of \SI{0.5}{\milli\meter}.

Next, the sample is attached to a commercial single-mode fiber ferrule and embedded in a superconducting RF microcoil with eight turns (Fig.~\ref{fig:PLEODMRHAMPIC}~(c)).
It provides a strong parallel magnetic driving field, that allows to achieve Rabi frequencies exceeding \SI{20}{\mega\hertz}.
An additional, macroscopic superconducting coil allows to apply DC magnetic bias fields along the crystal c-axis.
The fiber is used for both optical excitation and collection. The recorded optical signals are derived from photons emitted in the phonon sideband.
Further details on the experimental setup can be found in the Supplementary Material section I at \cite{SupplMat2024}.
%\textbf{First look}

In Fig.~\ref{fig:PLEODMRHAMPIC}~(a) we compare a photoluminescence-excitation (PLE) spectrum of the cylindrical microcrystal to a \SI{5}{\milli\meter} by \SI{5}{\milli\meter} bulk chip. 
The spectrum of the cylinder is significantly broadened and shifted towards lower energies compared to the bulk chip.
We ascribe this shift and modification of the optical spectrum to strain, likely induced by damage to the sidewalls of the cylinder during laser cutting.
Moderate transverse strain barely affects the optical transition frequency, so the measured spectral shift can be largely attributed to a longitudinal distortion.
Previous \textit{ab initio} and density functional theory (DFT) calculations indicate that the observed extent of the spectrum corresponds to a maximum $z$-strain of \(\sim 0.014 \,\%\) \cite{Tissot2024}.

In order to study the effect of strain on the spin transitions,  we measure the $|\downarrow,-7/2\rangle \leftrightarrow |\uparrow,-5/2\rangle$ clock transition at (\SI{29}{\milli\tesla}) using two different optical wavelengths. 
We acquire the optically detected magnetic resonance (ODMR) signal by applying a resonant laser pulse with a duration of \SI{500}{\micro\second}.
This pulse depletes the population of states with a corresponding optical resonance.
During  the laser pulse, a radio frequency pulse with a power of \SI{63}{\milli\watt} is applied for \SI{1}{\micro\second}. 
The ODMR signal is extracted from the increase in photoluminescence during the RF pulse, corresponding to a partial recovery of the population. A typical data set of the acquired photon counts can be seen in the Supplementary Material section II Fig.~2 at \cite{SupplMat2024}.
Two such measurements are shown in Fig.~\ref{fig:PLEODMRHAMPIC}~(b). The clock transition frequency is seen to decrease with increasing optical wavelength, from \SI{433.3 \pm 0.5}{\mega\hertz} to \SI{428.5 \pm 0.8}{\mega\hertz}. 
This behavior can be attributed to different strain acting on different parts of the PLE spectrum.
Assuming that the geometry can be chosen such that the component transforming like \(y\) vanishes, that the strain transforming like \(z\) correlates with a strain of similar magnitude which transforms like \(x\), and using the coupling constants calculated in \cite{Tissot2024}, we can estimate that the strain induced GS KD mixing angle is of the order of 
$\sim$
0.1
-
0.2
, depending on the particular strain elements at play.
Assuming further that the (thus far not experimentally determined) hyperfine parameters of the unstrained second GS KD and coupling the two unstrained GS KDs are \(\sim 100-200\,\)MHz results in a leading order shift of the clock transition frequency of \(\sim 0.4-6.5\,\)MHz using the perturbative energies \cite{SupplMat2024}, which is broadly in agreement with the data presented in Fig.~\ref{fig:PLEODMRHAMPIC}~(b).

Transverse strain is also predicted to lead to additional allowed transitions in the magnetic resonance spectrum (see Fig.~\ref{fig:PLEODMRHAMPIC}~(d)).
To observe these features, measurements of the RF response of the ensemble across a range of magnetic bias field strengths is recorded as described above. 
The spectra (and all following measurements) are recorded using an excitation wavelength of \SI{1279.12}{\nano\meter}. 
The resulting ODMR map is presented in Fig.~\ref{fig:map}.
Indeed, features beyond those expected for a strain-free sample (black lines in Fig.~\ref{fig:map} (a)) are clearly visible.
Small shifts between different measurement regions are visible which we attribute to hysteresis of magnetic material in the fiber ferrule assembly. Due to frequency-dependent heating of the sample, different powers are applied in different frequency ranges to ensure stable operation. 

The most notable differences are direct hyperfine transitions such as $|\downarrow,m\rangle \leftrightarrow |\downarrow,m\pm 1\rangle$ and $|\downarrow,m\rangle \leftrightarrow |\downarrow,m\pm 2\rangle$ at low frequencies up to a field of $\sim$\SI{40}{\milli\tesla}.
At higher frequencies, the recorded signal also significantly deviates from the strain free Hamiltonian.
The spectral features at low field are shifted and new transitions are observed close to the transitions expected for a strain free sample.

The high-resolution ODMR measurement provides strong support for the hypothesized, strain-induced modifications of the spin eigenbasis.
It further allows us to estimate the hyperfine parameters of the system using equation \ref{eq:Hg}.
We find modifications to the previously used hyperfine terms $a_{xx}$, $a_{yy}$ and $a_{zz}$. Additionally, $|a_{xz}|>0$ is required for the appearance of the lower branch of the direct hyperfine transitions, while it is predicted to be zero by group theory for intact $C_{3V}$ symmetry.
We use the following hyperfine parameters
(see Supplementary Material at \cite{SupplMat2024} section IV for details) to calculate the magnetic transition frequencies for this sample and plot it in Fig. \ref{fig:map} (b): $a_{xx} = \SI{176.5}{\mega\hertz}$, $a_{yy} = \SI{-148.5}{\mega\hertz}$, $a_{zz} = \SI{-230}{\mega\hertz}$, and $a_{zx} = \SI{5}{\mega\hertz}$.
These parameters are in broad agreement with the GS KD mixing angle as estimated above.
In the figure, the width of the plotted lines is proportional to the calculated signal strength of the individual transitions. The inclusion of strain in the Hamiltonian leads to good agreement between the calculated and measured frequencies and strengths of the transitions in the ODMR spectrum.

\begin{figure}
    %\centering
    \includegraphics[scale=1,page=2]{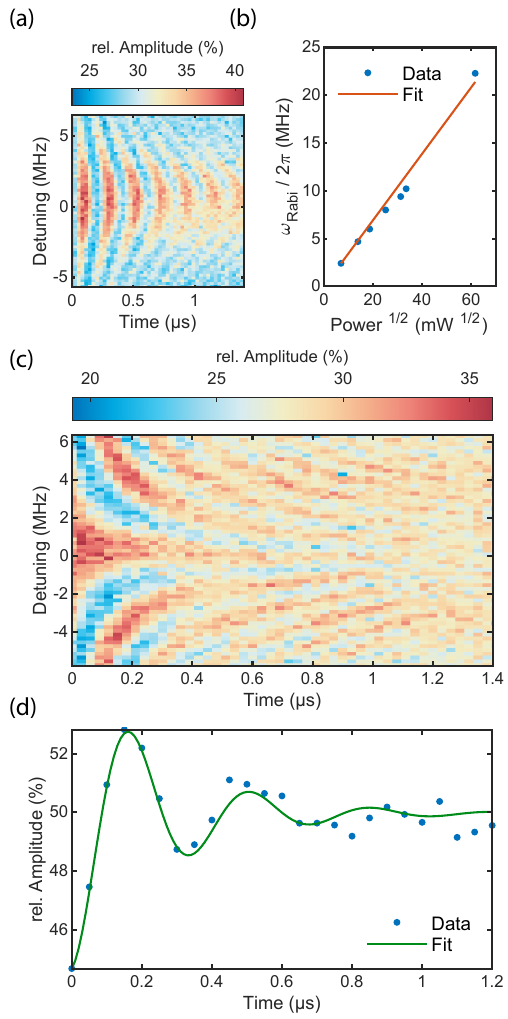}
    \caption{\label{fig:rabi} (a) Chevron pattern of Rabi oscillations vs. detuning.
    (b) Typical square root scaling of the Rabi frequency as function of the applied RF power with fastest observed Rabi frequency of \SI{22.3 \pm 0.3}{\mega\hertz}.
    The red line is a fit to the square root scaling.
    (c) Ramsey spectroscopy as function of the detuning to the transition center frequency.
    Data sets (a) and (c) were taken at the center frequency \SI{563}{\mega\hertz} and a field of approximately \SI{18.5}{\milli\tesla}. 
    (d) Rabi oscillations of the direct hyperfine transition $|\downarrow,m=-3/2\rangle \leftrightarrow |\downarrow,m=1/2\rangle$ driven with an RF frequency of \SI{81.5}{\mega\hertz} at approximately \SI{17.5}{\milli\tesla}.
    A fit to the data gives a Rabi frequency of \SI{2.9 \pm 0.13}{\mega\hertz} and a characteristic damping time scale of $T_2^{Rabi}=$\SI{0.260 \pm 0.03}{\micro\second}.
    }
\end{figure}

These modifications also indicate that in the new eigenbasis, most originally allowed transitions are accompanied by new, near-resonant transitions between other level pairs, as can be seen in Fig.~\ref{fig:map} (b).
In the absence of pure-state initialization, these additional resonances make it challenging to observe coherent rotations of the spin.
In order to characterize the coherence properties of the ensemble, we therefore select the clock transition with the resonance frequency of \SI{563}{\mega\hertz} at a field of $\sim$\SI{18.5}{\milli\tesla}, which corresponds to the $|\downarrow,m=-5/2\rangle \leftrightarrow |\uparrow,m=-3/2\rangle$ transition.
Rabi oscillations of this transition for different detunings are shown in Fig.~\ref{fig:rabi} (a).
A Fourier transform presented in the Supplementary Material at \cite{SupplMat2024} Fig.~3 
shows hints of a second resonance shifted by \SI{1.7 \pm 0.14}{\mega\hertz}, but seemingly with the same Rabi frequency $\Omega$. The presence of this unexpected feature is confirmed in Ramsey measurements (see below). 

The amplitude reduction of the Rabi oscillations near resonance yields a damping time $T_2^{Rabi}=$\SI{0.78\pm0.37}{\micro\second}.
However, this timescale is likely shortened by the presence of multiple sub-ensembles.
These may, at least in part, be related to hyperfine interactions with lattice nuclei, which have been observed and studied in detail elsewhere \cite{Hendriks2022}.
The Rabi frequency of the oscillations follows the expected square-root dependence on the applied RF power (Fig.~\ref{fig:rabi} (b)). The superconducting microcoil allows to reach $\pi$-pulse durations shorter than \SI{25}{\nano\second}, competitive with those achieved using nano-constrictions in coplanar waveguides, albeit on a far larger crystal volume \cite{Stas2022}. From the geometry of the superconducting microcoil, we find a drive strength conversion factor of \SI{6.3\pm0.3}{\mega\hertz/\milli\tesla}. This value is in good agreement with our theoretical prediction for the gyromagnetic moment of this transition, $|\gamma|/2\pi=$\SI{6.1}{\mega\hertz/\milli\tesla}
(see Supplementary Material section V at \cite{SupplMat2024} for further details).

The spectral properties of the studied transition are scrutinized further using a Ramsey measurement consisting of two $\pi/2$ pulses with a duration of \SI{53}{\nano\second}, separated by a variable time delay (Fig.~\ref{fig:rabi} (b)). Once again, the spectrum of the Ramsey oscillations reveals a second transition
(Supplementary Material at \cite{SupplMat2024} Fig.~3) 
at a detuning of \SI{1.7 \pm 0.14}{MHz}. We extract a dephasing time of $T_2^*=$\SI{0.41\pm0.17}{\micro\second} for the ensemble.
The additional transition is not predicted by our strain model. However, strain-induced transitions are, for the most part, significantly weaker than the originally allowed features.
Since the Rabi frequencies of the two features are indistinguishable in our measurements, we hypothesize that the second transition is caused by hyperfine coupling to $^{29}$Si nuclei in the SiC host, as observed in previous reports \cite{Hendriks2022}.

Finally, we perform the first coherent driving of direct hyperfine transitions enabled by strain. 
We select the $|\downarrow,m=-3/2\rangle \leftrightarrow |\downarrow,m=1/2\rangle$ transition at \SI{81.5}{\mega\hertz} and a magnetic field of \SI{17.5}{\milli\tesla}. The acquired signal together with a fit to the data is presented in Fig.~\ref{fig:rabi} (d).
Driving with an RF power of \SI{630}{\milli\watt} results in a Rabi frequency of \SI{2.9 \pm 0.13}{\mega\hertz}, and a characteristic damping time scale of $T_2^{Rabi}=$\SI{0.260 \pm 0.03}{\micro\second} is observed.

This measurement confirms that, by introducing strain in the system, it becomes possible to coherently control direct hyperfine transitions in a convenient experimental geometry and with high rotation frequencies. This control constitutes a key step on the path to fault-tolerant encoding of quantum information in the qudit manifold \cite{Lim2023}.

In our investigation of the 4H SiC V$^{4+} \alpha$ defect we observed several features that were not explicable with the models used in \cite{Wolfowicz2020,Tissot2021hyper,Tissot2021spin,Hendriks2022,Tissot2022,Astner2024}.
Our detailed spectroscopic analysis of the system, however, shows close agreement with a modified spin Hamiltonian which includes the effects of strain. 
While the changes to the PLE of our sample are most likely introduced by a $z$-strain distribution throughout the sample, direct hyperfine transitions are observed all across the broadened spectrum, suggesting that transverse strain is present in the entire ensemble.
In fact, hints of strain acting on TM defects in SiC can already be seen in \cite{Wolfowicz2020} and \cite{Hendriks2022}.
However, the degree of strain varies in dependence of the geometrical shape and intrinsic lattice defects of the SiC crystal.
Thus, to measure the vanadium defect in a completely strain free environment will likely require a single isolated defect with strain control.

  In this modified spin basis, we have shown the adressability of a wide range of transitions using parallel magnetic driving fields and coherent spin manipulations of both the electronic and hyperfine manifold.
The insights garnered in this work lend additional support to a recently proposed strain engineering technique which may raise the operational temperature of V$^{4+}$ to liquid-helium temperatures \cite{Ahn2024}. 
Further work will be required to extend the theoretical framework in order to include coupling to Si and C nuclear spins.
Controlled and calibrated application of stress will allow to quantify the strain coupling constants for the ground state spin manifold,  providing the basis for quantitative sensing of strain in SiC materials and devices \cite{Falk2014,Trotta2012,Ziss2017}.
Initialization of the spin into a pure state will enable complete control over the qudit state space for the creation of a long-lived quantum memory \cite{Adambukulam2024, Tissot2022, Senkalla2024,Lim2024}.
Overall, the presented measurements on 4H SiC V$^{4+}$ underpin the favorable features of this impurity for applications in quantum computing and communication.

\section*{Data Availability Statement}
The data sets that support the plots within this paper are available from the corresponding author upon reasonable request.

\section*{Acknowledgments}
This research was funded by the European Union under the Horizon 2020 Research and Innovation Program under grant agreement No. 86721 (QuanTELCO) and the Austrian Research Promotion Agency project FFG FO999914034 (SPQV).
CzechNanoLab project LM2023051 funded by MEYS CR is gratefully acknowledged for the financial support of the sample fabrication at CEITEC Nano Research Infrastructure.

\section*{Author Contributions}
P.K. and T.A. contributed equally to this work. B. T. and G. B. devised the theoretical framework and provided the theoretical description. The manuscript was written by all authors. G.B. and M.T. led the work.

\nocite{*}

%\bibliographystyle{apsrev4-2}
%\bibliography{quipcite}
%apsrev4-2.bst 2019-01-14 (MD) hand-edited version of apsrev4-1.bst
%Control: key (0)
%Control: author (72) initials jnrlst
%Control: editor formatted (1) identically to author
%Control: production of article title (-1) disabled
%Control: page (0) single
%Control: year (1) truncated
%Control: production of eprint (0) enabled
%

\end{document}